# Remarkable enhancement in crystalline perfection, Second Harmonic Generation Efficiency, Optical Transparency and laser damage threshold in KDP crystals by L-threonine doping

S. K. Kushwaha[1], Mohd. Shakir[1,2], K. K. Maurya[1], A. L. Shah[3], M.A. Wahab[2] and G. Bhagavannarayana[1]

[1]*Materials Characterization Division, National Physical Laboratory, Council of Scientific and Industrial Research, New Delhi-110 012, India*

[2]*Crystal growth Laboratory, Dept. of physics, Jamia Millia Islamia, New Delhi 110 025, India*

[3]*Solid State Laser Division, LASTEC, Metcalfe House, New Delhi – 110 054, India*

Effect of L-threonine (LT) doping on crystalline perfection, second harmonic generation (SHG) efficiency, optical transparency and laser damage threshold (LDT) in potassium dihydrogen phosphate (KDP) crystals grown by slow evaporation solution technique (SEST) has been investigated. The influence of doping on growth rate and morphology of the grown crystals has also been studied. Powder X-ray diffraction data confirms the crystal structure of KDP and shows a systematic variation in intensity of diffraction peaks in correlation with morphology due to varying LT concentration. No extra phase formation was observed which is further confirmed by Fourier Transform (FT) Raman studies. High-resolution X-ray diffraction curves indicate that crystalline perfection has been improved to a great extent at low concentrations with a maximum perfection at 1 mol% doping. At higher concentrations (5 to 10 mol%), it is slightly reduced due to excess incorporation of dopants at the interstitial sites of the crystalline matrix. LDT has been increased considerably with increase in doping concentration, whereas SHG efficiency was found to be maximum at 1 mol% in correlation with crystalline. The optical transparency for doped crystals has been increased as compared to that of pure KDP with a maximum value at 1 mol% doping.

## I. INTRODUCTION

The modern era of information technology with fast and high data storage capacity, data retrieving, processing and transmission has demanded the search for new nonlinear optical (NLO) materials with unique optical properties.[1] NLO materials in their single crystal form have wide applications in high-energy lasers for inertial confinement fusion research[2], color display, electro-optic switches, frequency conversion etc.[3] Hence, there is a great demand to synthesize new NLO materials and grow their single crystals. Most of the anisotropic physical, optical and electrical properties of the single crystals get deteriorated or completely diminished when these are not in the single domain crystal or having the defects like structural grain boundaries.[4,5] In parallel to the invention of new NLO materials, it is also important to modify the physical, optical and electrical properties of these materials either by adding functional groups[6] or incorporation of dopants[7,8] for tailor made applications. In the presence of dopants growth promoting factors like growth rate[9] and many of the useful physical properties like optical transparency[10,11], second harmonic generation (SHG) efficiency[8], laser damage threshold (LDT) etc. get enhance. The dopants or additives also influence the crystalline perfection which may in turn influence the physical properties depending on the degree of doping and as per the accommodating capability of the host crystal. Except a few investigations8 such studies are rare to found in the literature, which provide the interesting correlation between the properties and crystal quality. Potassium di-hydrogen phosphate (KDP) is a well known inorganic NLO material, having good ferroelectric, piezoelectric and electro-optic properties.[12,13] KDP favors for its crystal growth in bulk size, therefore its bulk size crystals have been easily grown by different techniques[14-16] suitable for device applications. From the inertial confinement fusion application point of view there are many studies regarding laser induced damage (LID) or laser damage threshold (LDT) of KDP crystals with significant improvements or modifications both in experimental and methodology aspects towards the improvement of their measurements. [17-20] In the present investigation influence of L-thronine doping in KDP crystals on growth morphology, SHG and LDT properties have been investigated. Pure and L-threonine (LT) doped single crystals of KDP were grown by the conventional slow evaporation solution technique (SEST). Powder X-ray diffraction (PXRD) has been used to study the effect of dopant concentration on the structure of KDP. FT-Raman studies have been carried out in the wave number range of 100-3500 cm-1. Influence of LT doping on SHG and LDT has been studied using neodymium-doped yttrium aluminium garnet; $Nd:Y_3Al_5O_{12}$ (Nd:YAG) laser source. The optical absorbance spectra for pure and doped specimens have been recorded in the complete ultraviolet-visible (UV-Vis) (200-800 nm) range and analyzed.

## II. EXPERIMENTAL

Commercially available 99% pure KDP salt was recrystallized for further purification and then used for single crystal growth by SEST method. Pure and LT added (1, 5 and 10 mol% concentration) KDP saturated solutions were prepared according to the solubility in double distilled water at 308 K with constant stirring. The solutions were filtered in beakers and covered with the perforated plastic sheet and then housed in a constant temperature bath at 308 K (±0.01 K) for crystal growth. Following the metastable zone-width for KDP[21], the temperature of constant temperature bath was reduced up to 303 K at the rate of 0.5 K per day and then kept constant. The well faceted with different morphology and transparent single crystals of pure and LT doped KDP were harvested from the mother solutions after a span of 20 days. The crystals with varied morphology due to LT doping are shown in Fig. 1. The change in the morphology may be due to different interactions of LT with different surfaces of KDP and variation of the concentration of the functional groups[22-23] of LT. Addition of the dopants also enhance the metastable zone width which also in turn leads to increase in the growth rate along particular direction(s),[21,24] whereas in other directions it decreases or remains same. The (100) planes consist alternatively positively $K^+$ and negatively $H_2PO_4^-$ groups and hence these are neutral and therefore inactive for the charge of the nearby liquid layers, whereas (101) planes are positively charged due to the termination of $K^+$ at the surface. The added impurity ions block the prismatic (100) faces and therefore lead to faster growth rate for (101) pyramidal planes.[9]

To reveal the influence of LT doping on the structure of KDP crystals the powdered specimens of pure and doped crystals were subjected to a PW1830 Philips analytical X-ray diffractometer with nickel filtered $CuK_\alpha$ radiation (35 kV, 30mA). The powders of the homogeneous particle size were obtained by crushing and filtering by ~ 25 micron sieve. All the specimens were scanned for the angular range of 10 to 70 degree of 2theta with the scan rate of 0.01 degree/sec, keeping all the experimental settings constant. Perkin Elmer GX 2000 FT-RAMAN spectrometer in the range of 100–3500 $cm^{-1}$ with the resolution of 1 $cm^{-1}$ has been used to carry out the functional and vibrational studies of pure and L-threonine doped KDP crystals. Good quality single crystals of the appropriate sizes were lapped and optically polished. The laser beam was made to incident normally on the prepared surfaces and scattered intensity was collected in the back scattering mode. The temperature of the room was maintained at 300 K to avoid any thermal influence on the vibrational data. The crystalline perfection of the undoped and LT doped KDP single crystals was characterized by HRXRD by employing a multicrystal X-ray diffractometer developed[25] at National Physical Laboratory (NPL), India. The well-collimated and monochromated MoK$\alpha_1$ beam obtained from the three monochromator Si crystals set in dispersive (+,-,-) configuration has been used as the exploring X-ray beam. The specimen crystal is aligned in the (+,-,-,+) configuration. Due to dispersive configuration, though the lattice constant of the monochromator crystal(s) and the specimen are different, the unwanted dispersion broadening in the diffraction curve (DC) of the specimen crystal is insignificant. The specimen can be rotated about the vertical axis, which is perpendicular to the plane of diffraction. The DC was recorded by the so-called $\omega$ scan wherein the detector was kept at the same angular position $2\theta_B$ with wide opening for its slit. Before recording the diffraction curve, to remove the non-crystallized solute atoms remained on the surface of the crystal and also to ensure the surface planarity, the specimen was first lapped and chemically etched in a non preferential etchant of water and acetone mixture in 1:2 volume ratio. This process also ensures the removal of possible complexating surface layers.[26] The relative SHG efficiency of pure and LT doped crystals was measured by Kurtz powder method by using KDP as standard reference. The homogeneous powder prepared by crushing and filtering through 25 MICS test sieve was densely filled in a glass micro-capillary of 1 mm inner bore and subjected to the focused beam of Nd:YAG [Spectra Physics (DCR-II)] laser at Indian Institute of Science (IISc), Bangalore. The second harmonic radiation (532 nm) generated by the randomly oriented microcrystals was focused by a lens and detected by a photomultiplier tube after filtration of the incident or fundamental radiation (1064 nm). The experimental settings were kept same for all the specimens to analyze the relative influence of LT doping on SHG efficiency of KDP. To find out the influence of LT doping on LDT of KDP single crystals, a Q-switched Nd:YAG pulsed laser with pulse energy 40 mJ and pulse width of 20 ns has been used. The laser beam with 2.5 mrad was focused to 0.16 mm spot size using a glass lens having 100 mm focal length. The neutral density (ND) filters have been used to vary the power density of laser beam from 0.50 to 11.00 GW/cm$^2$, and the output from these filters was delivered to the specimen crystal positioned at the near focus of converging lens. A pyro-detector calibrated at 1064 nm has been used to measure the energy of the laser beam. Before subjecting to the focused laser beam, the natural surfaces of the grown crystals were properly cleaned well polished and dried at 60 ºC to get rid from the probable dust and moisture. The UV-Vis. absorption spectra were recorded in the wavelength range of 200-800 nm on the Perkin Elmer Lambda 35 spectrophotometer with a resolution of 1 nm wavelength. The single crystal specimens of pure and LT doped crystals of the same thickness were used.

## III. RESULTS

### A. Powder XRD analysis

The PXRD patterns of pure and L-threonine doped KDP crystals are shown in Fig. 2 which contains all the original peaks of KDP. The spectra show that phase of KDP did not change with LT doping at all the concentrations. Due to doping, no new phase has been observed except a minute variation (which is well within the resolution limit) in the lattice parameters. For 1 mol%, except a few like (312), the intensity of all the peaks increases, which indicates the improved crystallinity. When the concentration of LT increases, the peak positions of the spectra remain constant whereas the intensity of the peak corresponding to (200) planes increases continuously and attained maximum intensity for the specimen doped with 10 mol% LT. High intensity of this peak is due to the enlargement of surface of (100) planes (Figure 1) due to faster growth of planes normal to it, in the presence of dopant molecules. This is because (100) planes of KDP have larger tendency[22,27] to expand as compared to that of (101). The impurity molecules and cations get easily adsorbed on the neutral (100) surfaces as compared to that of the (101) planes with +ve charge due to termination of the surfaces with $K^+$ and lead to faster growth of these planes.[9] Adsorbed dopant molecules on (100) surface also create the hindrance for the impurity molecules to enter into the crystalline matrix and lead to better crystalline perfection. However, the possibility of some portion of the LT dopants at substitutional sites cannot be ruled out.[28] The details of the effects of LT doping on the crystalline perfection is discussed in the forthcoming HRXRD analysis section.

### B. FT-Raman studies

The Raman spectra of pure and doped specimens were recorded for (100) planes at room temperature in the wavenumber range of 100 to 3500 $cm^{-1}$. However, the important portion of the spectra in the range of 100 to 1200 $cm^{-1}$ is shown in Fig. 3. The spectra (a), (b), (c) and (d) respectively of pure, 1.0, 5.0 and 10.0 mol% LT doped KDP single crystal specimens contain the internal modes of vibrations[28-29] of $H_2PO_4^-$ in $KH_2PO_4$ at 914 $cm^{-1}$ ($v_1$), 530 $cm^{-1}$ ($v_2$), 476 $cm^{-1}$ ($v_3$), 390 $cm^{-1}$ ($v_4$) and 357 $cm^{-1}$ ($v_5$). The peaks with very small intensity at around 213 and 166

cm$^{-1}$ are corresponding to the lattice vibrations of crystals through the absorption or emission of optical phonons. It is clearly observed from the figure that doping of L-threonine did not influence the internal vibrational modes of crystals as there is no shift or broadness of the main peak at 914 cm$^{-1}$ corresponding to the asymmetric stretching vibration of $H_2PO_4^-$ anions at all concentrations. The same nature of all spectra confirms no deviation in tetragonal phase and also reveals the absence of any additional phase with LT doping at any concentration which is in tune with the powder XRD.

## C. High-resolution X-ray diffraction

The curve in Fig. 4 shows the high-resolution diffraction curve (DC) recorded for undoped KDP specimen using (200) diffracting planes in symmetrical Bragg geometry with MoK$\alpha_1$ radiation. The curve is reasonably sharp having full width at half maximum (FWHM) of 7 arc s as expected for a nearly perfect crystal from the plane wave dynamical theory of X-ray diffraction.[30] Absence of additional peaks shows that the specimen crystal does not contain any internal structural grain boundaries[4] and indicates that the crystalline perfection is quite good.

Curves (a), (b) and (c) in Fig. 5 are DCs recorded respectively for the 1, 5 and 10 mol% LT doped KDP specimens under the identical conditions as that of Fig. 4. As seen in the figure, the DCs at all the doping levels contain single and sharp peaks with FWHM values 3.5, 3.8 and 8 arc s respectively for curves (a), (b) and (c). When these values are compared with that of the DC of Fig. 4, the following interesting observations can be revealed. Due to 1 mol% doping, FWHM decreased from 7 to 3.5 arc s. And also, when we compare the DCs of 1 mol% doped specimen and undoped specimen [Fig. 5(a) and Fig. 4] one can clearly see that the intensity along both the wings of the curve also reduced to a great extent after doping. This feature indicates that the dopant LT seems to hold the residual impurities present in the solution originated from the raw material and do not allow them to enter into the crystalline matrix. When the doping concentration is increased to 5 mol%, no much difference can be seen and similar dopant effect persists except a small increase of 0.3 arc s in FWHM with respect to that of curve (a) of Fig. 5. This gives an indication that the dopants are entering into the interstitial spaces of the lattice. This is clearer when we increase the dopant concentration up to 10 mol% as seen in curve (c), FWHM is increased considerably to 8 arc s. However, this value is still within the limits expected for a nearly perfect crystal.

It is interesting to see the asymmetry of the DC. For a particular angular deviation ($\Delta\theta$) of glancing angle with respect to the peak position, the scattered intensity is much more in the positive direction in comparison to that of the negative direction. This feature clearly indicates that the crystal contains predominantly interstitial type of defects than that of vacancy defects. This can be well understood by the fact that due to interstitial defects, the lattice around these defects undergo compressive stress[8] and the lattice parameter d (interplanar spacing) decreases leading to more scattered (also known as diffuse X-ray scattering) intensity at slightly higher Bragg angles ($\theta_B$) as d and $\sin\theta_B$ are inversely proportional to each other in the Bragg equation ($2d \sin\theta_B = n\lambda$; n and $\lambda$ being the order of reflection and wavelength respectively which are fixed). More details may be obtained from the study of high-resolution diffuse X-ray scattering measurements[25,26] which is however not the main focus of the present investigation. It is worth to mention here that the observed scattering due to interstitial defects is of short range order as the strain due to such minute defects is limited to the very defect core and the long range order could not be expected and hence the change in the lattice parameter of the crystal is not possible. It may be mentioned here that the minute information like the asymmetry in the DC could be possible as in the present sample only because of the high-resolution of the multicrystal X-ray diffractometer used in the present investigation.

Increase of FWHM without any splitting of DC along with the asymmetry of the DC [curve (c)] corresponding to the peak position as explained above clearly shows the incorporation of LT in the crystalline matrix of KDP. HRXRD studies elucidate the fact that LT not only restricts the entry of impurities in the crystal lattice, which resembles the effect of ethylenediamine tetra-acetic acid organic additive in zinc tris-thiourea sulphate (ZTS) and ammonium dihydrogen phosphate (ADP) crystals[31], but also entering into the crystalline matrix up to some extent interstitially when the doping concentration in the solution is high as observed in our recent study on KCl and oxalic acid doped ADP crystals.[8]

### D. SHG efficiency

The SHG for pure and doped crystals was confirmed by output green radiation from single crystals during the LDT measurements. The measured relative SHG efficiency of undoped and doped crystals is plotted with LT concentration as shown in Fig. 6(a). The curve in the figure shows strong nonlinear dependence of SHG on the doping concentration. It attained sharp maximum value (1.32 times to that of pure KDP) for 1 mol% and then slowly decreased. It may be mentioned here that the crystalline perfection as observed from HRXRD studies is sharply increased at 1 mol% LT doping and then slowly decreased as the concentration increased. These results indicate the strong correlation of SHG on crystalline perfection which in turn depends on doping concentration. In our recent studies[31] also, a direct effect of crystalline perfection on SHG efficiency of ZTS and ADP crystals has been observed. At higher concentrations, particularly when the size of the dopant is large, SHG efficiency does not increase, rather it may decrease due to deterioration of crystalline perfection as observed in ADP crystals.[6] In the present investigation also at higher concentration, crystalline perfection slightly decreased followed by the reduction of SHG efficiency. However, even at 10 mol%, the SHG efficiency is still a little bit higher than that of the pure KDP as no much deterioration has been taken place for the crystalline perfection which may otherwise possible by the formation of structural grain boundaries at very higher concentrations.[8] In LiNbO$_3$, the maximum SHG has been attained by the reduction of Li vacancies[5] which are otherwise common to observe in these crystals due to evaporation of Li during crystal growth. Similarly, our earlier studies[5] on LiNbO$_3$ crystals which were annealed (followed by poling) with slow heating and cooling rates, crystalline perfection was enhanced, which in turn lead to enhancement of the piezoelectric coefficient d$_{33}$ to a great extent up to 23 pC/N from the reported value of 17 pC/N for samples which were not subjected to such annealing. These examples also confirm a direct bearing of crystalline perfection on the many important physical properties.

### E.  Laser damage threshold

The laser induced breakdown in the crystals caused by various physical processes such as electron avalanche, multiphoton absorption and photoionization for the transparent materials whereas in case of high absorbing materials, the damage threshold is mainly due to the temperature rise, which leads to strain-induced fracture.[32-33] It also depends upon the specific properties of material, pulse width, and wavelength of laser used. For the long-pulse regime $\tau > 100$ ps, the damage process occur mainly by the rate of thermal conduction through the atomic lattice and for the short-pulse regime $\tau \leq 10$ ps, the optical breakdown is a nonthermal process and various nonlinear ionization mechanisms (multiphoton, avalanche multiplication, and tunneling) become important.[33] In the present investigation, 20 ns pulsed laser has been used and therefore thermal effects are prominent. LDT values for (100) planes of pure and doped crystals were recorded when the clear visible spot occurred on surface with audible sound. As given in the experimental details, to get the bulk LDT values, the beam was focused inside the crystal, but close to the surface with a predetermined spot size of diameter 0.16 mm in air. As the spot size is very small and it may not be the same inside the crystal due to change in the refractive index, there could be some error in the value of the actual spot size inside the crystal which in turn lead to error in the calculation of energy density. In view of this fact, we have given the relative LDT values by taking the LDT value of pure KDP as one unit. These values are plotted with the doping concentration and shown in Fig. 6(b) which indicates that the relative LDT value first increased sharply with 1 mol% doping and increased gradually at higher concentrations. This feature indicates that LT doping has a strong influence on the enhancement in LDT irrespective of crystalline perfection.

### F. UV-Vis absorbance

The recorded UV-Vis. absorbance spectra of pure and doped KDP single crystals in the wavelength region 200 to 800 nm is shown in Fig. 7. As seen in the figure, pure crystals have maximum absorbance in the entire spectral range in comparison with the doped specimens. For 1 mol%, absorbance has been reduced to a great extent and becomes almost transparent for the entire UV-Vis optical range. This may be due to the maximum crystalline perfection of KDP at this concentration which also may be one of the reasons for this specimen to become most SHG efficient. For 5 and 10 mol% LT doped specimens, the absorbance is slightly higher than that of 1 mol% doped specimen, but still lesser than that of pure KDP. Similar increase in the UV transparency of KDP has also been reported due to KCl doping.[21]

### IV. CONCLUSION

Present investigation revealed the following conclusions. L-threonine doping has a strong effect on growth rate and morphology in KDP crystals. Doping lead to faster growth rate for (101) pyramidal planes leading to growth of crystals with larger surface area for (100) planes. The crystalline phase of the KDP does not change even at the higher concentrations of doping whereas an interesting variation in the peak intensities of PXRD patterns was found due to changes in morphology. All the modes of vibrations for the tetragonal phase of pure KDP have been observed in the FT-Raman spectra. No peak shift has been noticed at all the concentrations of dopant. Low LT doping concentration enhanced crystalline perfection to a great extent whereas at higher concentrations it is slightly reduced due to incorporation of LT in the crystalline matrix. SHG efficiency also varied in the same manner as that of crystalline perfection. The relative laser damage threshold was found to be enhanced with LT doping at all concentrations. The optical transparency for doped crystals has been increased as compared to that of pure KDP with a maximum value at 1 mol% doping where the crystalline perfection is maximum. These studies elucidated the fact that L-threonine doping lead to a great enhancement in crystalline perfection, SHG, LDT and optical transparency with a strong correlation between the crystalline perfection and SHG as well as optical transparency.

**ACKNOWLEDGEMENTS**

Authors acknowledge Director NPL, for constant encouragement in carrying out the present investigation. They also acknowledge the Director, Laser Science and Technology (LASTEC) for giving permission to carry out the LDT studies at LASTEC. Kushwaha acknowledges Council of Scientific and Industrial Research (CSIR) for providing the Senior Research Fellowship.

**REFERENCES**

[1]D. J. Williams, Angew Chem., Int. Ed. Engl. **23**, 690 (2003).

[2]N. Zaitseva and L. Carman, Prog. Cryst. Growth Charact. **43**, 1 (2001).

[3]J. Badan, R. Hierle, A. Perigaud, and J. Zyss, *NLO properties of Organic Molecules and Polymeric Materials* (American Chemical Society Symposium Series 233; American Chemical Society: Washington, DC, 1993).

[4]G. Bhagavannarayana, R. V. Ananthamurthy, G. C. Budakoti, B. Kumar and K. S. Bartwal, J. Appl. Cryst. **38**, 768 (2005).

[5]G. Bhagavannarayana, G. C. Budakoti, K. K. Maurya, and B. Kumar, J. Cryst. Growth **282**, 394 (2005).

[6]Sweta Moitra and Tanusree Kar, Opt. Mater. **30**, 508 (2007).

[7]F. Zernike and J. E. Midwinter, *Applied nonlinear optics* (Wiley, New York, 1973).

[8]G. Bhagavannarayana, S. Parthiban and S. Meenakshisundaram, Cryst. Growth Des. **8**, 446 (2008).

[9]S. A. Vries de, P. Goedtkindt, S. L. Bennett, W. J. Huisman, M. J. Zwanenburg, D. M. Smilgies, J. J. De Yoreo, W. J. P. Enckevort van, P. Bennema, and E. Vlieg, Phys. Rev. Lett. **80**, 2229 (1998).

[10]S. Kar, R. Bhatt, K. S. Bartwal and V. K. Wadhawan, Cryst. Res. Technol. **39**, 230 (2004).

[11]E. Winkler, P. Etchegoin, A. Fainstein, and C. Fainstein, Phys. Rev. B **61**, 15756 (2000).

[12]K. B. R. Varma, K. V. Ramanaiah, and K. Veerabhadra Rao, Bull. Mater. Sci. **5**, 39 (1983).

[13]W. Jamroz, K. Karniewicz, and J. Stachowiak, Sov. J. Quantum Electron. **9**, 803 (1979).

[14]N. Balamurugan and P. Ramasamy, Cryst. Growth Des. **6**, 1642 (2006).

[15]S. Balamurugan and P. Ramasamy, Mater. Chem. Phys. **112**, 1 (2008).

[16]S. Balamurugan and P. Ramasamy, Spectrochimica Acta A, **71**, 1979 (2009).

[17]C. W. Carr, M. D. Feit, M. C. Nostrand and J. J. Adams, Meas. Sci. Technol. **17**, 1958 (2006).

[18]L. Lamaignere, S. Bouillet, R. Courchinoux, T. Donval, M. Josse, J. C. Poncetta, and H. Bercegol, Rev. Sci. Instr. **78**, 103105 (2007).

[19]I. M. Pritula, V. I. Salo and M. I. Kolybaeva, Inorg. Mater. 37, 184 (2001).

[20]H. Yoshida, T. Jitsuno, H. Fujita, M. Nakatsuka, M. Yoshimura, T. Sasaki, K. Yoshida, Appl. Phys. B **70**, 195 (2000).

[21]Guohui Li, Liping Xue, Genbo Su, Zhengdong Li, Xinxin Zhuang and Youping He, Cryst. Res. Technol. **40**, 867 (2005).

[22]Dongli Xu and Dongfeng Xue, Physica B **370**, 84 (2005).

[23]Y. L. Geng, D. Xu, Y. L. Wang, W. Du, H. Y. Liu, G. H. Zhang, X. Q. Wang, and D. L. Sun, J. Cryst. Growth **273**, 624 (2005).

[24]P. V. Dhanaraj, K. Santheep Mathew, and N. P. Rajesh, J. Cryst. Growth **310**, 2532 (2008).

[25]K. Lal and G. Bhagavannarayana, J. Appl. Cryst. **22**, 209 (1989).

[26]G. Bhagavannarayana, S. Parthiban and S. P. Meenakshisundarm, J. Appl. Cryst. **39**, 784 (2006).

[27]C. M. R. Remédios, W. Paraguassu, P. T. C. Freire, J. Mendes-Filho, J. M. Sasaki, and F. E. A. Melo, Phys. Rev. B **72**, 014121 (2005).

[28]G. W. Lu, and X. Sun, Cryst. Res. Technol. **37**, 93 (2002).

[29]B. W. Batterman and H. Cole, Rev. Mod. Phys. **36**, 681 (1964).

[30]P. H. Dederichs, J. Phys. **F**: Met. Phys. **3,** 471 (1973).

[31] M. H. Li, Y. H. Xu, R. Wang, X. H. Zhen, and C. Z. Zhao, Cryst. Res. Technol. **36,** 191 (2001).

[32] W. L. Smith, Opt. Eng. **17**, 489 (1959).

[33] B. C. Stuart, M. D. Feit, A. M. Rubenchik, B. M. Shore, and M. D. Perry, Phys. Rev. Lett. **74**, 2248 (1995)

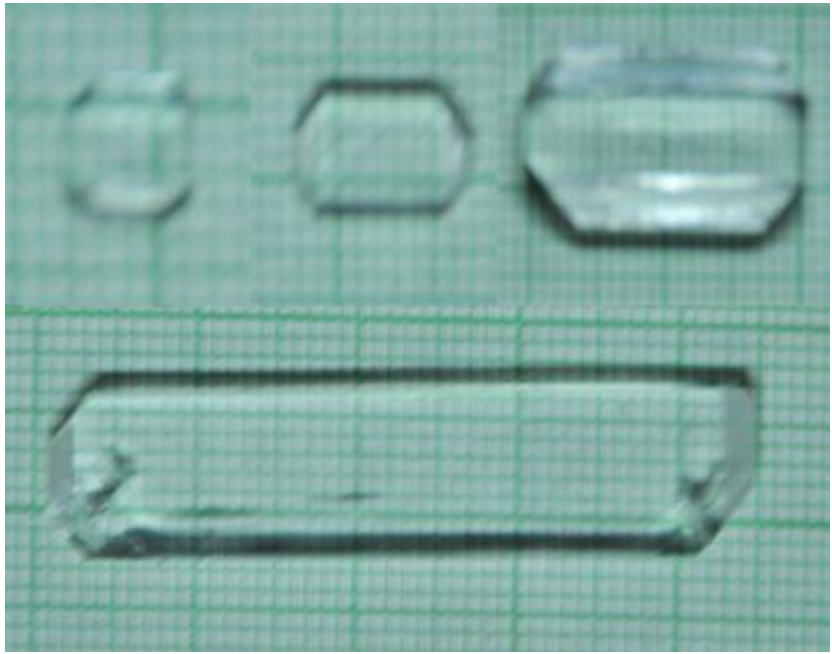

FIG. 1. SEST grown (a) undoped, (b) 1.0, (c) 5.0 and (d) 10 mol% LT doped KDP single crystals

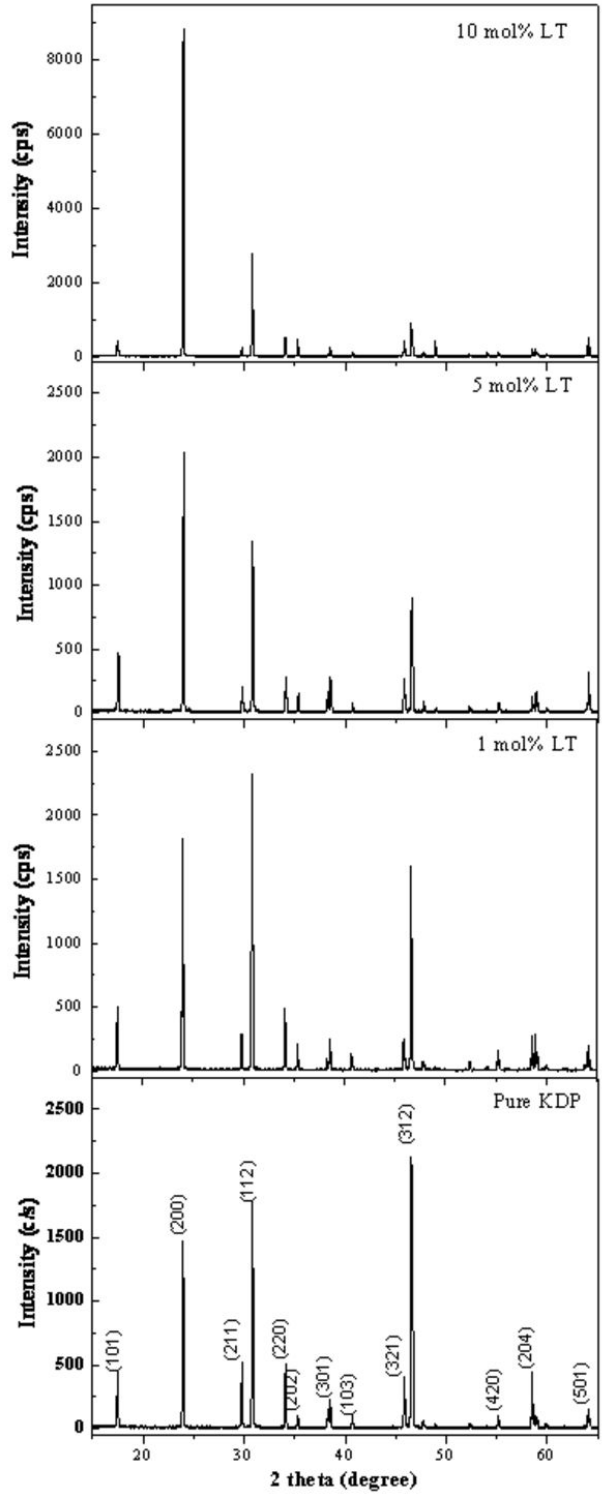

FIG. 2. PXRD pattern of pure and LT doped KDP with different concentrations

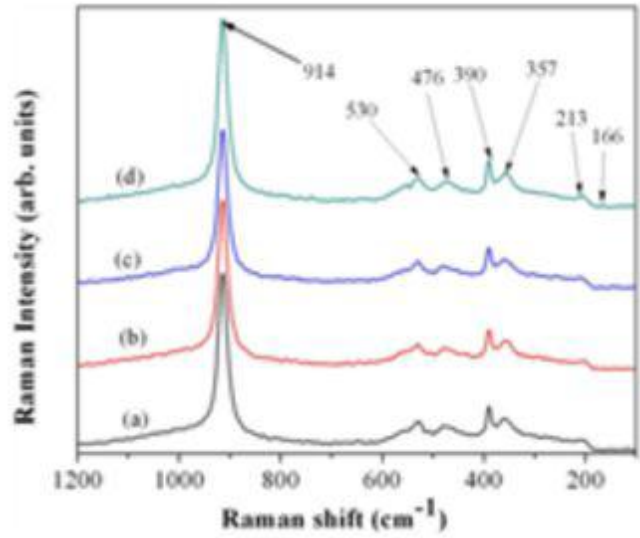

FIG. 3. FT-Raman spectra (a), (b), (c) and (d) for pure, 1, 5 and 10 mol% LT doped KDP single crystals respectively

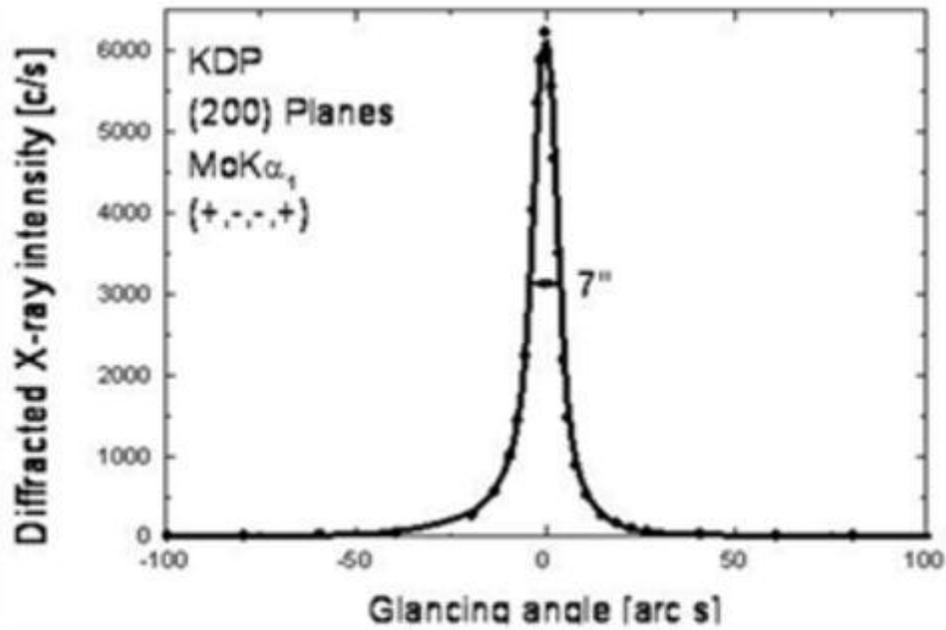

FIG. 4. HRXRD curve for pure KDP single crystal recorded for (200) diffraction planes

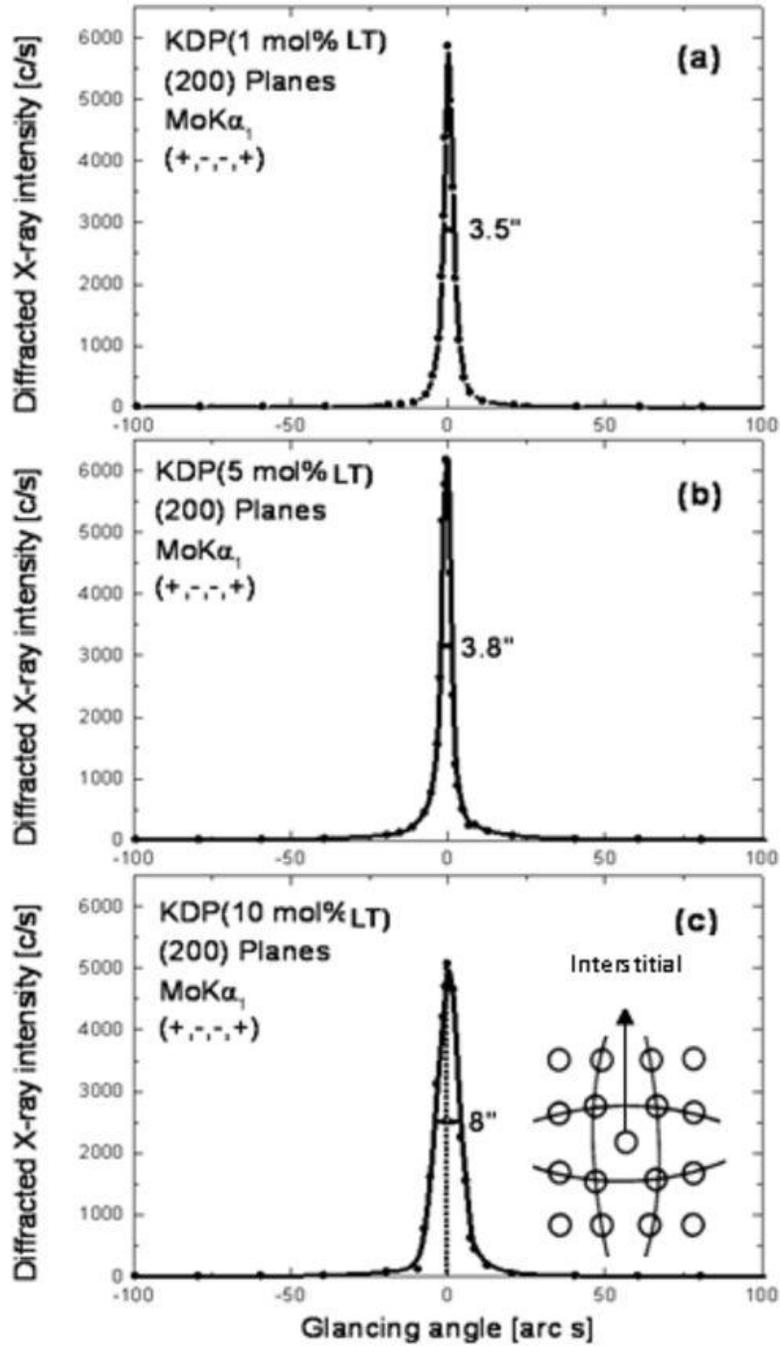

FIG. 5. HRXRD curves of (a) 1 mol%, (b) 5 mol% and (c) 10 mol% LT doped KDP single crystals recorded for (200) diffraction planes. The inset in (c) depicts the compressive stress in the lattice around the defect core

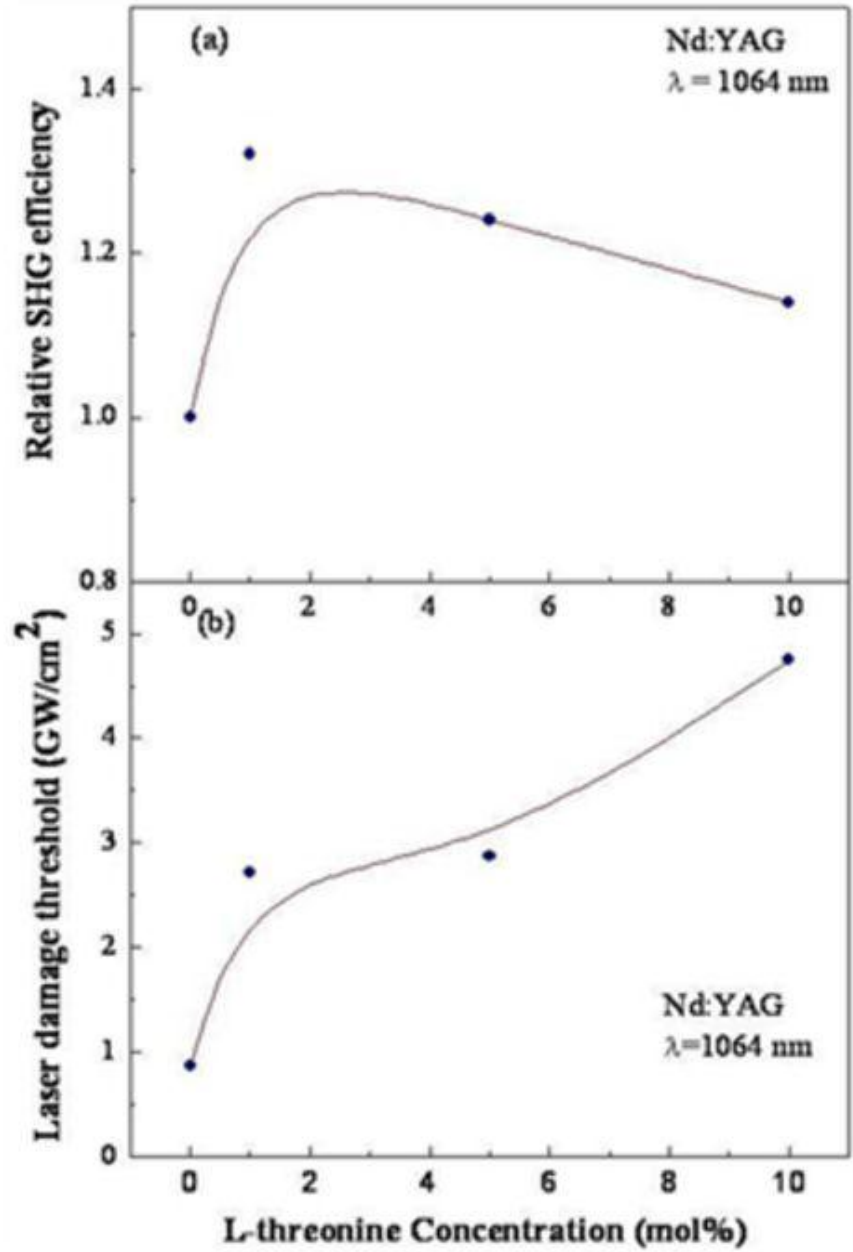

FIG. 6. Relative (a) SHG efficiency and (b) laser damage threshold of KDP crystals with different doping concentrations of LT

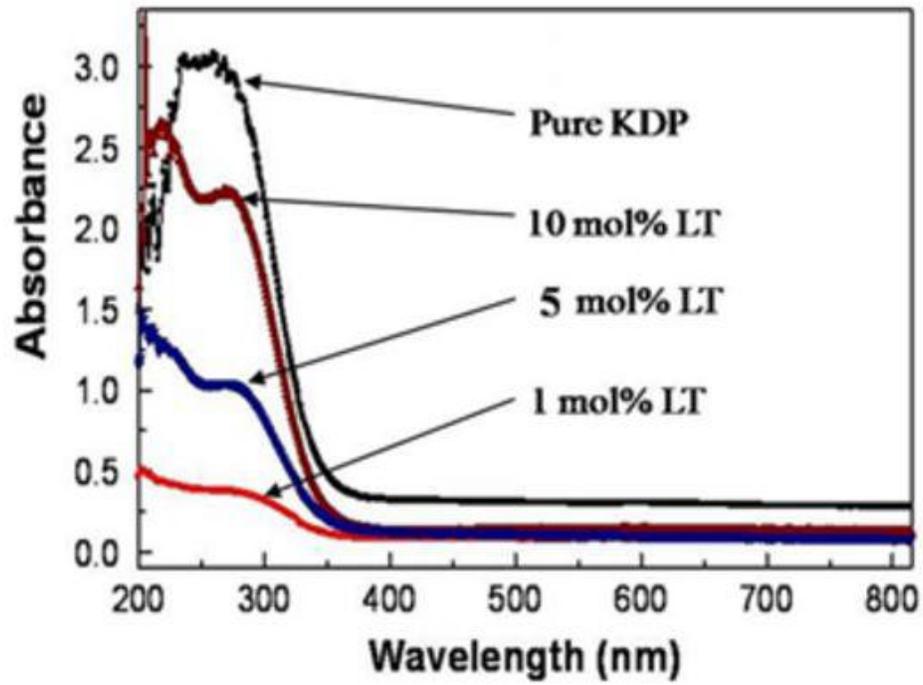

FIG. 7. The optical absorbance of pure and LT doped KDP single crystals